# Social Influence Distorts Ratings in Online Interfaces

Marina Kontalexi[*]
marinak@sam.sdu.dk
University of Southern Denmark
Odense, Denmark

Alexandros Gelastopoulos[†]
alexandros.gelastopoulos@iast.fr
Institute for Advanced Study in
Toulouse, France

Pantelis P. Analytis
pantelis@sam.sdu.dk
University of Southern Denmark
Odense, Denmark

## Abstract

Theoretical work on sequential choice and large-scale experiments in online ranking and voting systems has demonstrated that social influence can have a drastic impact on social and technological systems. Yet, the effect of social influence on online rating systems remains understudied and the few existing contributions suggest that online ratings would self-correct given enough users. Here, we propose a new framework for studying the effect of social influence on online ratings. We start from the assumption that people are influenced linearly by the observed average rating, but postulate that their propensity to be influenced varies. When the weight people assign to the observed average depends only on their own latent rating, the resulting system is linear, but the long-term rating may substantially deviate from the true mean rating. When the weight people put on the observed average depends on both their own latent rating and the observed average rating, the resulting system is non-linear, and may support multiple equilibria, suggesting that ratings might be path-dependent and deviations dramatic. Our results highlight potential limitations in crowdsourced information aggregation and can inform the design of more robust online rating systems.

## 1 Introduction

One of the promises of the Web has been the prospect to readily crowdsource information from diverse groups of people in the form of ratings. User or consumer ratings are omnipresent today and constitute a key part of the business model for some of the most recognizable companies of the Internet era (e.g., Amazon, Airbnb, Tripadvisor, Vivino, etc.). Explicit feedback in the form of ratings is, in turn, a crucial component for designing effective ranking and recommendation algorithms. Although accumulating information online is easier than ever before, people are likely influenced by the ratings that they see before they evaluate a product on their own [9, 25]. Can social influence distort the long-term ratings of products? If so, how large should we expect these distortions to be?

Theoretical work on sequential (or simultaneous) choice has shown that social influence can have dramatic consequences, as people might be led astray by the already established popularity of an option and eventually settle on options of inferior quality [1, 2, 4, 6, 12]. Large scale empirical studies on ranking and upvoting/downvoting interfaces [16, 17] have also demonstrated that social influence can substantially alter the social dynamics and the long-term outcomes of online systems, influencing the evaluations and/or popularity of the curated content. Surprisingly, there is comparatively little work assessing the implications of social influence on online rating interfaces, where people evaluate items on a specific rating scale (e.g., 1-10 stars)—potentially the most common sequential interaction systems on the web [7, 15, 22, 24]. Much of this work suggests that, given enough interactions, the ratings of products in these systems would self-correct and converge to the same average rating that they would receive if people rated independently [23, 24].

In this paper, we put forward a new framework for studying the effect of social influence on online rating systems that leverages results from the theory of stochastic dynamical systems [3, 13]. As in earlier work on this topic, we start from the empirically grounded assumption that people are linearly influenced by the average they observe before they contribute their own rating for a product [20, 25]. We show that when people's latent independent rating for products and their propensity to be influenced are correlated, the resulting system is linear, yet social influence can lead to substantial distortions in long-term online ratings. Further, we demonstrate that when the weight people assign to the observed rating depends on both their own latent rating and the observed rating, the system becomes nonlinear, and in some settings it may even be path-dependent, implying that the long-term average rating could be determined by the decisions of a few early raters. Our approach is open-ended and can be applied to evaluate other social influence models, other interaction settings (e.g., estimation problems), as well as to propose design interventions that could lead to better aggregate-level outcomes.

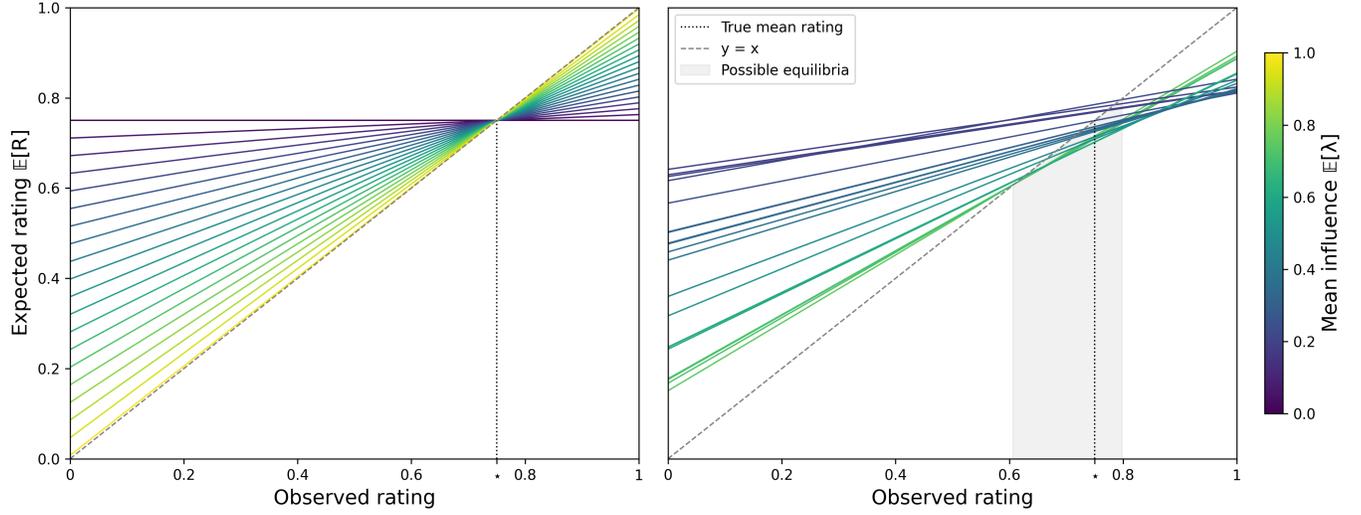

Figure 1: Social influence curves for the cases where the agent's influence parameter $\lambda_i$ and latent rating $r_i$ are independent (left) or correlated (right). The point where a line intersects the diagonal $y = x$ is the unique equilibrium of the system; as $N \to \infty$, the average rating $\overline{R}_N$ converges to that point. Left: When $\lambda_i$ and $r_i$ are independent, the equilibrium is not affected by social influence and it is always equal to the true mean rating. The different lines are obtained by setting $\mathbb{E}[r_i] = 0.75$ and varying $\mathbb{E}[\lambda_i]$ in the range $[0, 1]$ (see eq. (2)). Right: When $\lambda_i$ and $r_i$ are correlated, the equilibrium depends on the specifics of their distributions (eq. (3)), implying that social influence has permanent effects on the average ratings. The shaded gray region shows the interval over which the equilibrium ranges. The different lines are obtained from eq. (2) by varying the distribution of $\lambda_i$, while keeping $\mathbb{E}[r_i]$ constant at 0.75 (see Methods).

## 2 Theoretical model and results

### 2.1 Social influence model

We study a model of social influence, where a very large population of individuals sequentially rate an item, with continuous evaluations in $[0, 1]$. When it is the $i$-th agent's turn to rate, she is also presented with the average value of all previous ratings. We assume that the agent has an independent opinion $r_i$ upon the quality of the item, which we refer to as her *latent* rating. Her *expressed* rating $R_i$ is given by

$$R_i = \lambda_i \cdot x + (1 - \lambda_i) \cdot r_i, \quad (1)$$

where $x = \overline{R}_{i-1} = \frac{1}{i-1} \sum_{j=1}^{i-1} R_j$ is the average expressed rating of all previous users and $\lambda_i \in [0, 1]$ is called the influence parameter. The terms average expressed rating and observed rating are used interchangeably throughout the paper. According to this definition, the higher the value of $\lambda_i$, the stronger the social influence on agent $i$'s rating. We assume that the latent ratings $r_i$ are drawn i.i.d. from a fixed distribution. We call $\mu = \mathbb{E}[r_i]$ the *true mean rating* of the item; this is the average rating that the item would receive if all users expressed their independent opinions. We are interested in whether the average expressed rating $\overline{R}_N = \frac{1}{N} \sum_{i=1}^{N} R_i$ converges to the true mean rating $\mu$ as more and more users rate the item ($N \to \infty$).

### 2.2 Main results

Previous literature assumes that the influence parameter $\lambda_i$ is homogeneous across the population. We study the case in which the influence parameter $\lambda_i$ not only differs across agents, but it might also correlate with the user's latent rating $r_i$. We first look at the special case where $\lambda_i$ may correlate with $r_i$, but does not depend on the observed rating $x$ (= average expressed rating of previous users), and then we generalize further to allow $\lambda_i$ to depend on both $r_i$ and the observed rating $x$.

#### 2.2.1 Influence parameter does not depend on observed rating $x$.

Assuming that $\lambda_i$ does not depend on $x$ and taking conditional expectation given $x$ in eq. (1), we obtain[1]

$$\mathbb{E}[R_i \mid x] = \mathbb{E}[\lambda_i] \cdot x + \mathbb{E}[(1 - \lambda_i) r_i], \quad (2)$$

that is, the expected expressed rating $R_i$ of the $i$-th user, conditional on the current average expressed rating $x = \overline{R}_{i-1}$, is a linear function of $x$. Figure 1 plots this function when $\lambda_i$ and $r_i$ are either (a) independent or (b) correlated. Consistent with previous work on social influence in binary choice settings, we call the function $f(x) := \mathbb{E}[R_i \mid x]$ the *influence curve* of the system [11]. This function expresses the expected value of the rating that a randomly chosen user will give to an item, assuming that upon entering the platform they see $x$ as that item's average rating (based on previous users' evaluations).

The form of $f(x)$ can tell us what will happen to the average rating in the long-term. Specifically, note that if for some value of $x$ we have $f(x) > x$, this means that the ratings received from future users will on average be higher than the current average $x$. This will result in the average rating increasing as more users enter the platform. In contrast, if $f(x) < x$, then future users will on average evaluate the item with lower ratings, thus the average rating of the item will tend to decrease. Thus, the only way for the average rating to stabilize is if the expected rating from future users is *equal* to its current rating, that is $f(x) = x$. Such values are called equilibria of the system and they correspond to the points where the influence curve crosses the $45^o$ line $y = x$. In other words, as the number of users becomes large, the average rating *must* converge to one of the points where the graph of the influence curve $f(x)$ crosses the line $y = x$. For a rigorous mathematical argument see [5, Theorems 1 & 2].

---
[1]We assume that $r_i$ is also independent of $x$, since $r_i$ represents a latent rating that exists independently of external influences.

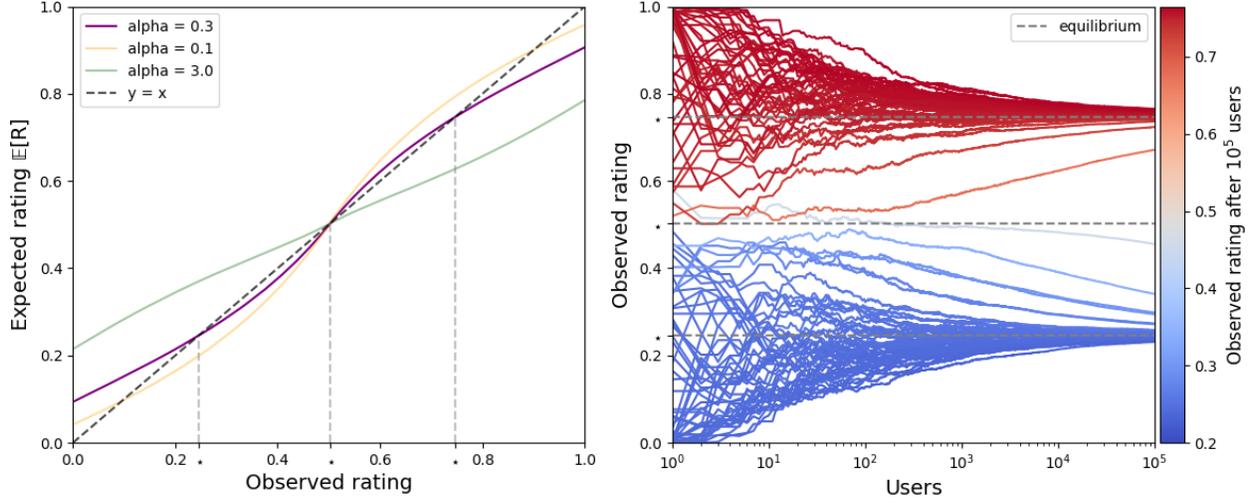

Figure 2: Left: Social influence curves when the influence parameter depends on both the latent rating $r_i$ and the current average rating $x$. Here, $\lambda_i$ is chosen so that, on average, it decreases with the distance of the agent's latent rating $r_i$ from the observed rating $x$ (see Methods). The different curves correspond to different distributions of the latent ratings $r_i$, specifically, $r_i \sim \text{Beta}(\alpha, \alpha)$, with smaller values of $\alpha$ corresponding to more polarized populations. In the cases $\alpha = 0.1$ and $\alpha = 0.3$, there are two stable equilibria (points where the curve downcrosses the diagonal, indicated for the case $\alpha = 0.3$ as the leftmost and rightmost vertical dashed gray lines), implying that the limit of the average rating $\overline{R}_N$ is ex ante unpredictable. For $\alpha = 3$, there is a single stable equilibrium at $x = 0.5$, implying that $\overline{R}_N \to 0.5$ necessarily. Right: Simulations with $10^4$ agents with $r_i$'s and $\lambda_i$'s as before, for $\alpha = 0.3$. The agent population remains fixed across simulations, but their order is randomized. The average rating $\overline{R}_n$ converges to either of the stable equilibria.

When the influence curve $f(x) := \mathbb{E}[R_i \mid x]$ is linear, it has a unique equilibrium (fig. 1). Its value can be found algebraically as the solution of the equation $\mathbb{E}[R_i \mid x] = x$. Denoting by $R^*$ this solution, we get from eq. (2) that $R^* = \mathbb{E}[\lambda_i] \cdot R^* + \mathbb{E}[(1 - \lambda_i)r_i]$, and solving for $R^*$,

$$R^* = \frac{\mathbb{E}[(1-\lambda_i)r_i]}{1 - \mathbb{E}[\lambda_i]} = \frac{\mathbb{E}[(1-\lambda_i)] \cdot \mathbb{E}[r_i] + \text{Cov}(1-\lambda_i, r_i)}{1 - \mathbb{E}[\lambda_i]}$$
$$= \mu - \frac{\text{Cov}(\lambda_i, r_i)}{1 - \mathbb{E}[\lambda_i]} \quad (3)$$
$$= \mu - \rho(\lambda_i, r_i) \cdot \frac{\sigma_{\lambda_i} \cdot \sigma_{r_i}}{1 - \mathbb{E}[\lambda_i]},$$

where in the second equality we have used the relation $\text{Cov}(X, Y) = \mathbb{E}[XY] - \mathbb{E}[X]\mathbb{E}[Y]$ and in the last one the relation $\text{Cov}(\lambda_i, r_i) = \rho(r_i, \lambda_i) \cdot \sigma_{r_i} \cdot \sigma_{\lambda_i}$, where $\sigma_{\lambda_i}$ and $\sigma_{r_i}$ denote the standard deviation of $\lambda_i$ and $r_i$, respectively, and $\rho(\lambda_i, r_i)$ their correlation coefficient. If there is correlation between $\lambda_i$ and $r_i$, then the empirical rating $R^*$ will deviate from the true mean rating $\mu$, leading to substantial systematic overestimation or underestimation (depending on the sign of $\rho(\lambda_i, r_i)$) of the true mean rating of a product (see fig. 1b).

However, if $\rho(\lambda_i, r_i) = 0$, i.e., $r_i$ and $\lambda_i$ are uncorrelated, then we get $R^* = \mu$, meaning that the average expressed rating $\overline{R}_N$ will converge to the true mean rating given enough users. Geometrically, this means that the influence curve crosses the $45^o$ line necessarily at $x = \mu$ (see also fig. 1a). This makes the system *self-correcting*, in the sense that any early discrepancies due to chance or due to systematic but short-lived attacks will not have a permanent effect on the rating of the item. This condition for self-correction is more general than in previous work that assumes that all individuals have the same propensity to be influenced regardless of their latent ratings. It shows that the system will self-correct even in a population with heterogeneous influence parameters, as long as these are uncorrelated with their latent ratings [23, 24].

*Remark*. In a setting where individuals rate multiple products (i.e. books, hotels, or wines) eq. (3) implies that the distortions in the ratings ($R^* - \mu$) can change the relative ranking of the products in terms of their ratings. Indeed, because the distributions of the latent ratings $r_i$ around their mean and of the influence parameters $\lambda_i$ might differ across items, the degree of distortion will generally also vary, and it can result in two items with $\mu_1 > \mu_2$ having average ratings at equilibrium $R_1^* < R_2^*$. For a system to be rank-preserving, (i.e., $R_1^* > R_2^*$ whenever $\mu_1 > \mu_2$), the quantity $\rho(\lambda_i, r_i) \cdot \frac{\sigma_{\lambda_i} \cdot \sigma_{r_i}}{1-\mathbb{E}[\lambda_i]}$ would have to be item-independent, which would hold in the uncorrelated case ($\rho(\lambda_i, r_i) = 0$), but rarely otherwise.

*2.2.2 The observed rating $x$ directly moderates influence.* In some cases, it is reasonable to assume that people's propensity to be influenced might be moderated not only by their latent rating $r_i$ itself, but also by the deviation of $r_i$ from the observed average rating $x$. For example, people might be less susceptible to be influenced by the average rating $x$ if it significantly deviates from their latent rating $r_i$, because in that case they would find the average rating to be an unreliable estimate of the true quality of the item. Inversely, it could also be the case that the influence parameter *increases* with the distance of their latent rating from $x$, because they might suspect that their personal experience was less representative of a typical use of the item [10].

We now extend some of the results of the previous section to the case where the distribution of $\lambda_i$ may depend on both $x$ and $r_i$. We maintain the assumption that the latent rating $r_i$ is independent of $x$.

Taking conditional expectations in eq. (1), conditioned on the average rating $x$, we get

$$\begin{aligned} \mathbb{E}[R_i \mid x] &= \mathbb{E}[\lambda_i x + (1 - \lambda_i) r_i \mid x] \\ &= \mathbb{E}[\lambda_i (x - r_i) \mid x] + \mathbb{E}[r_i \mid x] \\ &= \mathbb{E}[\lambda_i (x - r_i) \mid x] + \mu. \end{aligned} \quad (4)$$

Now $\mathbb{E}[R_i \mid x]$ is generally nonlinear as a function of $x$, so the equation $\mathbb{E}[R_i \mid x] = x$ may have multiple solutions, i.e., the graph of the influence curve $f(x) := \mathbb{E}[R_i \mid x]$ may cross the diagonal line $y = x$ multiple times (fig. 2, left). The points where $f$ crosses the diagonal *downwards* (i.e., points $x^*$ for which $f(x) > x$ to the left of $x^*$ and $f(x) < x$ to the right of $x^*$) are stable equilibria, in the sense that the average rating $\overline{R}_N$ may converge to any of those with positive probability [13]. In contrast, points where $f$ crosses the diagonal upwards are unstable equilibria, because small discrepancies from those tend to be amplified.

The existence of multiple stable equilibria implies that there are multiple possible values for the long-term average rating $\lim_{N \to \infty} \overline{R}_N$, and its value is inherently unpredictable ex ante, crucially depending on the ratings of a few initial users (fig. 2, right; see also [2]). When multiple stable equilibria exist, they are typically widely separated (fig. 3), highlighting a potentially dramatic impact of (possibly adversarial) early raters on the long-term average rating of an item. However, even in cases where there is a single equilibrium, similar to the scenario discussed in section 2.2.1, substantial distortion of the average rating relative to the true mean rating $\mu$ can still occur (see the part of the parameter space that supports only one equilibrium in fig. 3).

Because the existence of multiple stable equilibria can have significant effects on the ratings of the items, it is important to understand the conditions under which these equilibria may arise. The following proposition demonstrates that multiple equilibria cannot occur in the special case that people become on average less influenced as the distance of the observed average rating $x$ from their latent rating $r_i$ increases.

PROPOSITION 2.1. *Suppose that for any fixed $r_i$, $\mathbb{E}[\lambda_i \mid x, r_i]$ is continuous and increasing in $x$ for $x \leq r_i$ and continuous and decreasing in $x$ for $x \geq r_i$. Then, $\mathbb{E}[R_i \mid x] = x$ has a unique solution.*

PROOF. Conditioning on both $r_i$ and $x$ in eq. (1), we get

$$\begin{aligned} \mathbb{E}[R_i \mid x, r_i] &= \mathbb{E}[\lambda_i (x - r_i) \mid x, r_i] + \mathbb{E}[r_i \mid x, r_i] \\ &= \mathbb{E}[\lambda_i \mid x, r_i] \cdot (x - r_i) + r_i. \end{aligned} \quad (5)$$

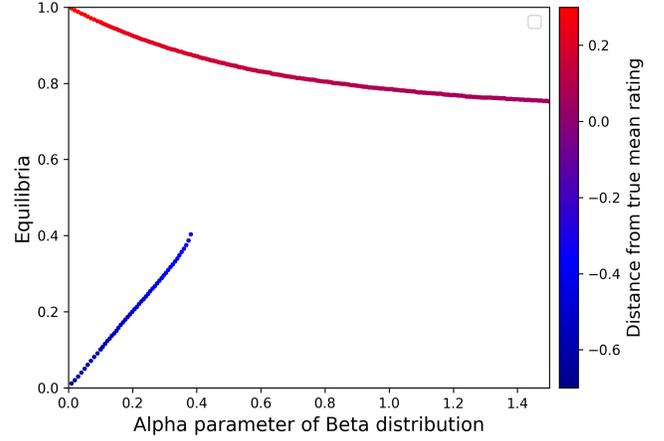

**Figure 3: Bifurcation diagram for $r_i \sim \text{Beta}(\alpha, \frac{\alpha}{0.7} - \alpha)$ (resulting in $\mu = \mathbb{E}[r_i] = 0.7$) and $\lambda_i$ as in fig. 2a, with $\alpha$ as the bifurcation parameter. For values of $\alpha$ smaller than $\approx 0.4$, there are two stable equilibria, while for larger values there is a single equilibrium but different from the true mean rating $\mu$. Lower $\alpha$ values correspond to increasing polarization in the population in terms of independent opinions about the item.**

From the assumptions on $\mathbb{E}[\lambda_i \mid x, r_i]$, we see that the above quantity is decreasing in $x$ in both intervals $[0, \mu]$ and $[\mu, 1]$, therefore it is decreasing in the entire interval $[0, 1]$ as well. Integrating over $r_i$, we get that this also holds for $\mathbb{E}[R_i \mid x]$. It follows that $\mathbb{E}[R_i \mid x] = x$ has at most one solution in $[0, 1]$. The fact that it has at least one solution follows from the intermediate value theorem applied to $g(x) := \mathbb{E}[R_i \mid x] - x$, since $g$ is continuous, $g(0) = \mathbb{E}[R_i \mid x = 0] - 0 \geq 0$, and $g(1) = \mathbb{E}[R_i \mid x = 1] - 1 \leq 0$. □

A unique equilibrium means that regardless of fluctuations in the rating during the early stages, the long-term average rating of the item will remain unaffected. Thus, when the condition of the aforementioned theorem is satisfied, early users cannot exert a significant influence on the eventual rating of a product. For example, a few early fraudulent ratings would not distort the emergent rating for the product. Only sustained efforts that effectively alter the underlying distribution of ratings, such as the consistent insertion of fraudulent ratings over time, would have a lasting impact.

## 3 Discussion and future directions

Producing information in the form of crowdsourced average ratings is an essential service provided by many online marketplaces (e.g., Amazon) and databases (e.g., Imdb). To what extent does the mean of the crowdsourced ratings accurately reflect the independent average opinions of the users, and what are the implications of social influence at the system level? In this study, we introduced a novel approach for studying this problem grounded in the theory of stochastic dynamical systems. This theoretical framework and the notions of self-correction and path-dependence have been previously used to study the implications of social influence in binary choice settings [2, 11, 21]. However, to the best of our knowledge, this is the first application of such a framework to scenarios where individuals adjust their responses along a continuous scale. We started with the empirically supported assumption that people are influenced by the current average rating in a linear manner and investigated the conditions under which the average of the crowd-sourced ratings self-corrects—i.e., converges to the true mean of the original rating distribution—or fails to do so. In cases where self-correction did not occur, the system could either converge to a distorted but unique equilibrium, or exhibit multiple possible equilibria, potentially diverging substantially from the true mean rating due to path-dependent dynamics.

Although we focused on a linear model of social influence, the framework we proposed can readily generalize to other models. The social influence curve is an aggregate level concept and can be used to analyze the social dynamics produced by alternative models of social influence or even combinations thereof. For instance, recent empirical work using data scraped from online interfaces suggests that, in some contexts, social influence curves may exhibit non-monotonic behavior [22, 26]. Such effects could arise if individuals strategically exaggerate their ratings in an attempt to shift the current average toward their own independent assessments [19, 26]. Future research should examine the dynamics produced by such a social influence model and the conditions under which they give rise to self-correcting or persistently distorted rating systems. Additionally, further research can examine whether simpler individual-level models can generate non-monotonic social influence curves [18]. Finally, we anticipate that our modeling framework will pave the way for experimental studies aimed at differentiating among competing individual-level models of social influence, thereby complementing insights derived from observational data on online rating platforms.

Going beyond online ratings, our framework can be used to study different settings altogether. For example, our model can be applied to investigate the implications of social influence in settings where interacting individuals estimate or forecast a continuous variable of interest, such as the growth rate of an economy, or the likelihood of an event [14]. These settings have the same structural properties with online rating systems—people express (and update) their estimates sequentially and their beliefs are likely to be influenced by the current collective estimate, as it tends to be more accurate than their own [8]. In fact, in some cases the linear weighted model might be the optimal belief update [20]. Such settings provide a natural domain for further applying the concepts that we put forward.

## Methods

For fig. 1b, we numerically estimate $\mathbb{E}[\lambda_i]$ and $\mathbb{E}[(1-\lambda_i)r_i]$ and then use eq. (2). For the numerical estimation, we first sample 10000 agents with $r_i \sim \text{Beta}(3, 1)$ (so that $\mathbb{E}[r_i] = 0.75$), and then draw $\lambda_i \sim \text{Beta}(3, 3/c(r_i, m) - 3)$, where $c(r_i, m) = \mathbb{E}[\lambda_i \mid r_i] := \frac{|r_i - m|}{max(m, 1-m)}$ and $m$ is a constant. The different lines were generated by sampling $m$ from a normal distribution with $\mu_m = 0.7$ and $\sigma_m = 0.2$, truncated so that $m \in [0, 1]$. In figs. 2 and 3 we used eq. (4) with $r_i$ as described in the figure captions and $\lambda_i \sim \text{Beta}(3, 3/c(r_i, x) - 3)$, where $x$ is the observed rating. Simulation code can be found by clicking here.

## Acknowledgments

We thank Ville Satopää and the members of the Strategic Organization Design group at the University of Southern Denmark for their valuable feedback. Pantelis P. Analytis, Marina Kontalexi and Alexandros Gelastopoulos are supported by the Sapere Aude research grant from the Independent Research Fund Denmark. Alexandros Gelastopoulos acknowledges support from ANR under grant ANR-17-EURE-0010 (Investissements d'Avenir program).